\documentclass[aps,prl,superscriptaddress,onecloumn,nobibnotes]{revtex4-1}

\usepackage{graphicx}
\usepackage{float}
\usepackage{array,url}
\usepackage{units}
\usepackage{epstopdf}
\usepackage{gensymb}

\begin{document}

\title{Cavity cooling of many atoms}
%\preprint{}

%\email[E-mail me at: ]{bugs@looney.com}
%\homepage[Visit: ]{http://looney.com/}
%\altaffiliation[Permanent address: ]{Warner Brothers}
\author{Mahdi Hosseini}
\altaffiliation[Current affiliation: ]{Birck Nanotechnology Center, School of Electrical and Computer Engineering, Purdue University, West Lafayette, Indiana 47906, USA}
\author{Yiheng Duan}
\author{Kristin M. Beck}
\altaffiliation[Current affiliation: ]{JQI and Department of Physics, University of Maryland, College Park, Maryland 20742, USA}
\affiliation{Department of Physics and Research Laboratory of Electronics, Massachusetts Institute of Technology, Cambridge, Massachusetts 02139, USA}
\author{Yu-Ting Chen}
\affiliation{Department of Physics and Research Laboratory of Electronics, Massachusetts Institute of Technology, Cambridge, Massachusetts 02139, USA}
\affiliation{Department of Physics, Harvard University, Cambridge, Massachusetts, 02138, USA}
\author{Vladan Vuleti\'c}
\affiliation{Department of Physics and Research Laboratory of Electronics, Massachusetts Institute of Technology, Cambridge, Massachusetts 02139, USA}
%\affil[2]{Department of Physics, Harvard University, Cambridge, Massachusetts, 02138, USA}
%\renewcommand\Authands{and}

%\author{Mahdi Hosseini, Yiheng Duan, Kristin M. Beck, Yu-Ting Chen and Vladan Vuleti\'c }
%\author{Kristin M. Beck}  
%\author{Yiheng Duan}
%\author{Vladan Vuleti\'c} \email{vuletic@mit.edu}
%\affiliation{Department of Physics and Research Laboratory of Electronics, Massachusetts Institute of Technology, Cambridge, Massachusetts 02139, USA}

\date{\today}

\begin{abstract}
We demonstrate cavity cooling of all motional degrees of freedom of an atomic ensemble using light that is far detuned from the atomic transitions by several gigahertz. 
The cooling is achieved by cavity-induced frequency-dependent asymmetric enhancement of the atomic emission spectrum, thereby extracting thermal kinetic energy from the atomic system.
Within $100 ~\mathrm{ms}$, the atomic temperature is reduced from $200 ~\mu\mathrm{K}$ to $10 ~\mu\mathrm{K}$, where the final temperature is mainly limited by the linewidth of the cavity. In principle, 
 the technique can be applied to molecules and atoms with complex internal energy structure. 
\end{abstract}
% \pacs{}

\maketitle

%INTRODUCTION
 The coherent interaction of atoms with an electromagnetic mode of a high finesse optical resonator can be used to control the electromagnetic field~\cite{kimblePRL1992,kimble:PS1998, Fushman:Sci:2008, Rempe:Nature2016, Rauschenbeutel:nphot:2014,Brooks2012}, or to entangle the internal states of many atoms~\cite{
 McConnell:Nature2015,Hosten:Nature2016,JThompson2011}. Moreover, the strong light-matter interaction provided by the optical resonator can be employed to control
 and cool the external degrees of freedom of atoms or other particles~\cite{Horak:PRL1997, VV:PRL2000,Rempe2006coolingPRA, Holland:PRL2016, VV:pra2001, Maunz:Nature2004, Ike:PRL2009, Monika:prl2011, Hemmerich:science, Chan:PRL2003,asenbaum2013cavity,millen2015cavity,kiesel2013cavity}, as well as massive oscillators~\cite{Marquardt2009,Metzger2004,Corbitt2007,Gigan2006,Arcizet2006,Schliesser2006,Thompson2008}. Notably, cooling with light far off resonant from any atomic or molecular transition becomes possible, as the sign of the velocity dependent force can be set by the frequency of the cavity, rather than that of the atomic transition~\cite{Horak:PRL1997, VV:PRL2000}. Cavity cooling uses the fact that the spectrum of the light scattered by a moving particle is broadened relative to the
  incident light, and contains both lower-frequency (Stokes) and higher-frequency (anti-Stokes) components, corresponding to an increase or decrease of the
  particle's kinetic energy, respectively. By tuning the cavity to the anti-Stokes sideband, it is then possible to cool moving objects via light scattering 
  into the cavity, in both the free-particle regime (cavity Doppler cooling) and the strong-confinement regime 
 (cavity sideband cooling)~\cite{VV:pra2001,Marquardt2007PRL}. To date, cavity cooling has been applied to single atoms~\cite{Maunz:Nature2004}, ions~\cite{Ike:PRL2009}, nanoscale particles~\cite{asenbaum2013cavity,millen2015cavity,kiesel2013cavity}, the center-of-mass mode of an atomic ensemble~\cite{ Monika:prl2011}, and nanomechanical oscillators~\cite{Marquardt2009,Metzger2004,Corbitt2007,Gigan2006,Arcizet2006,Schliesser2006,Thompson2008}. Moreover, a Bose-Einstein condensate has been transferred deterministically
 between two momentum states
via cavity scattering~\cite{Hemmerich:science}, and collective-emission-induced cooling of multilevel atoms in a low-finesse optical resonator has been
 observed~\cite{Chan:PRL2003}, although the exact mechanism for the latter has not been identified. \par
 
 In this Letter, we demonstrate simultaneous cavity cooling of all motional degrees of freedom in an ensemble containing a few hundred atoms. The cooling is performed at a large detuning of several gigahertz from atomic resonance. The maximal detuning is only limited by the low available laser power of a few $\mu\mathrm{W}$. The temperature is reduced by a factor of 20, and the phase space density increased by over two orders of magnitude, within $100 ~\mathrm{ms}$. The observations are well described by a simple single-particle model of cavity cooling~\cite{VV:pra2001}. The cooling rate is set by the photon scattering rate into the cavity at the given laser power and chosen detuning from atomic resonance, while the final temperature of $10(1)~\mu\mathrm{K}$ is limited by the cavity linewidth $\kappa$ ($\kappa=2\pi\times 160 ~\mathrm{kHz}$, $\hbar \kappa/k_B=7.6~\mu\mathrm{K}$ for our system).
  
 \begin{figure}[!th]
\centerline{\includegraphics[width=.4\columnwidth]{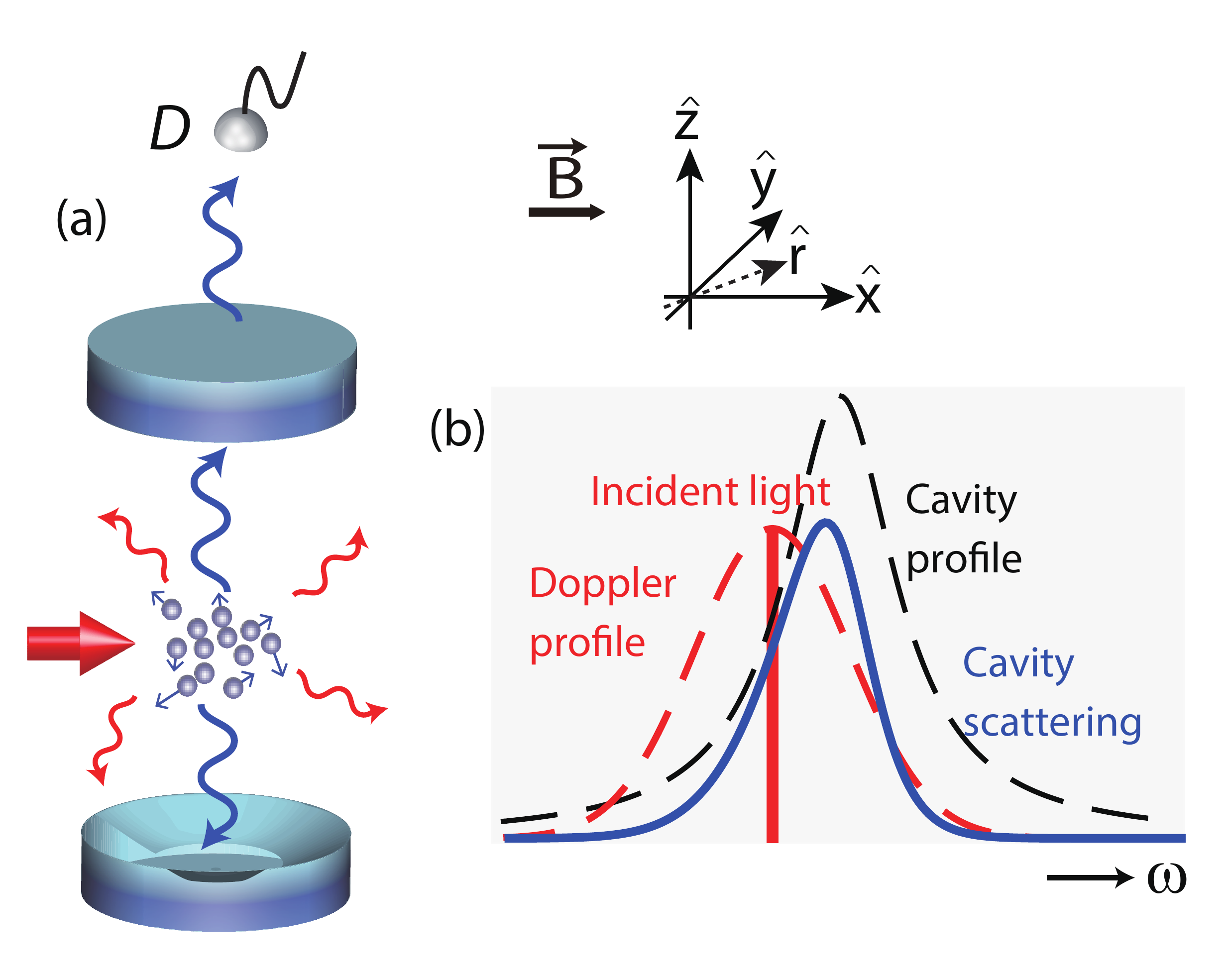}}
\caption{(a) Schematic representation of cavity cooling of an atomic ensemble. Laser light detuned from the atomic
 transitions by several hundred atomic linewidths, and slightly red-detuned from the cavity resonance frequency, illuminates the atoms
  from the side. (b) The blue-detuned part of the Doppler-broadened atomic emission spectrum (red dashed line) is enhanced by the cavity. Thus the light scattered into the cavity (blue solid line) has an average frequency that exceeds that of the incident light (red solid line), thereby extracting thermal energy from the atoms. The final temperature is set by the cavity linewidth. Light collected from the cavity on detector $D$ is used to measure the atomic temperature during cooling.}
   
   %(c) An interference measurement of the light emerging from the cavity 
   %shows spectral narrowing and a frequency shift associated with cavity cooling. The red and blue spectrum are for
   %atomic temperatures of 200 $\mu$K and 30 $\mu$K.   }
\label{fig1}
\end{figure}  

Our system consists of an ensemble of $\mathrm{^{133}Cs}$ atoms held within the TEM$_{00}$ mode of a high-finesse optical cavity that enhances both the cooling light (wavelength $\lambda_c=852 ~\mathrm{nm}$, finesse $\mathcal{F}_c=7.71(5)\times 10^4$) and the trapping light ($\lambda_t=937 ~\mathrm{nm}$, $\mathcal{F}_t=3.7(2) \times 10^2$). To load atoms into a small volume, so that we can achieve high total photon scattering rate even at large detuning from atomic resonance with limited laser power ({\raise.17ex\hbox{$\scriptstyle\mathtt{\sim}$}}$3~\mu$W), we initially load the atoms from a magneto-optical trap (MOT) into a single-beam dipole trap formed by a 937-nm trapping beam propagating normal to the cavity mode (along $\hat{x}$), and focused to a waist of $2 ~\mu$m. We then transfer atoms from this single-beam dipole trap to the intracavity standing-wave dipole trap. In this way, we create a small atomic cloud of about 200 atoms trapped primarily at two antinodes of the cavity standing-wave trap (along $\hat{z}$). The radial and axial trap vibrational frequencies are $\omega_{rad}/2\pi=3~\mathrm{kHz}$ and $\omega_{ax}/2\pi=350 ~\mathrm{kHz}$, respectively, and the initial peak atomic density is $n_0=4.5\times10^{12}~\mathrm{cm}^{-3}$. The typical temperature of the atomic ensemble in the cavity dipole trap prior to cavity cooling is $T_{i}$ ${\raise.17ex\hbox{$\scriptstyle\mathtt{\sim}$}}$ $200 ~\mu\mathrm{K}$. Since the upper hyperfine state manifold $F=4$ of the electronic ground state $6S_{1/2}$ exhibits an unusually large inelastic collision cross section that leads to fast atom loss at our atomic densities~\cite{Chin:PRL2000}, we continuously deplete the $F=4$ manifold using near-resonant $6S_{1/2},~F=4\to 6P_{3/2},~F'=4$ depumping light. The cavity cooling light is red-detuned by several GHz from the $F=3\to F'=2$ transition, propagating normal to the cavity axis (along the $\hat{x}$ direction, Fig.\ref{fig1}(a)), and focused to a waist of $10 ~\mu\mathrm{m}$ at the atom's location. It is also red-detuned by approximately a quarter of the cavity linewidth ($\vert \delta_i \vert/2\pi ~{\raise.17ex\hbox{$\scriptstyle\mathtt{\sim}$}} ~40 ~\mathrm{kHz}$) from the cavity resonance, such that the cavity enhances the blue-detuned part of the atomic Doppler emission spectrum, thereby reducing the thermal energy of the atoms in the cavity scattering process (Fig.\ref{fig1}(b)). 
 A magnetic field along the $\hat{x}$ axis, $B_x=3 ~\mathrm{G}$, sets the quantization
 axis. The far-detuned effective cooperativity of the cavity is given
 by the ratio of the single-photon Rabi frequency, $2g$, to atomic ($\Gamma$) and cavity ($\kappa$) energy decay rates, $\eta=4g^2/\kappa\Gamma$. Due to pointing fluctuations of the single-beam dipole trap, the cooperativity varies between the maximum value of 5, averaged over hyperfine atomic transitions, at the antinodes of the cavity standing wave and the minimum value of 0 exactly at the nodes. For the analysis of the data in this Letter, we use the averaged value of $\eta=2.5$. The cooperativity equals the ratio of the scattering rate into the resonant cavity to the scattering rate into free space~\cite{Haruka:AAMOP2011}. 

\begin{figure}[!th]
\centerline{\includegraphics[width=1.\columnwidth]{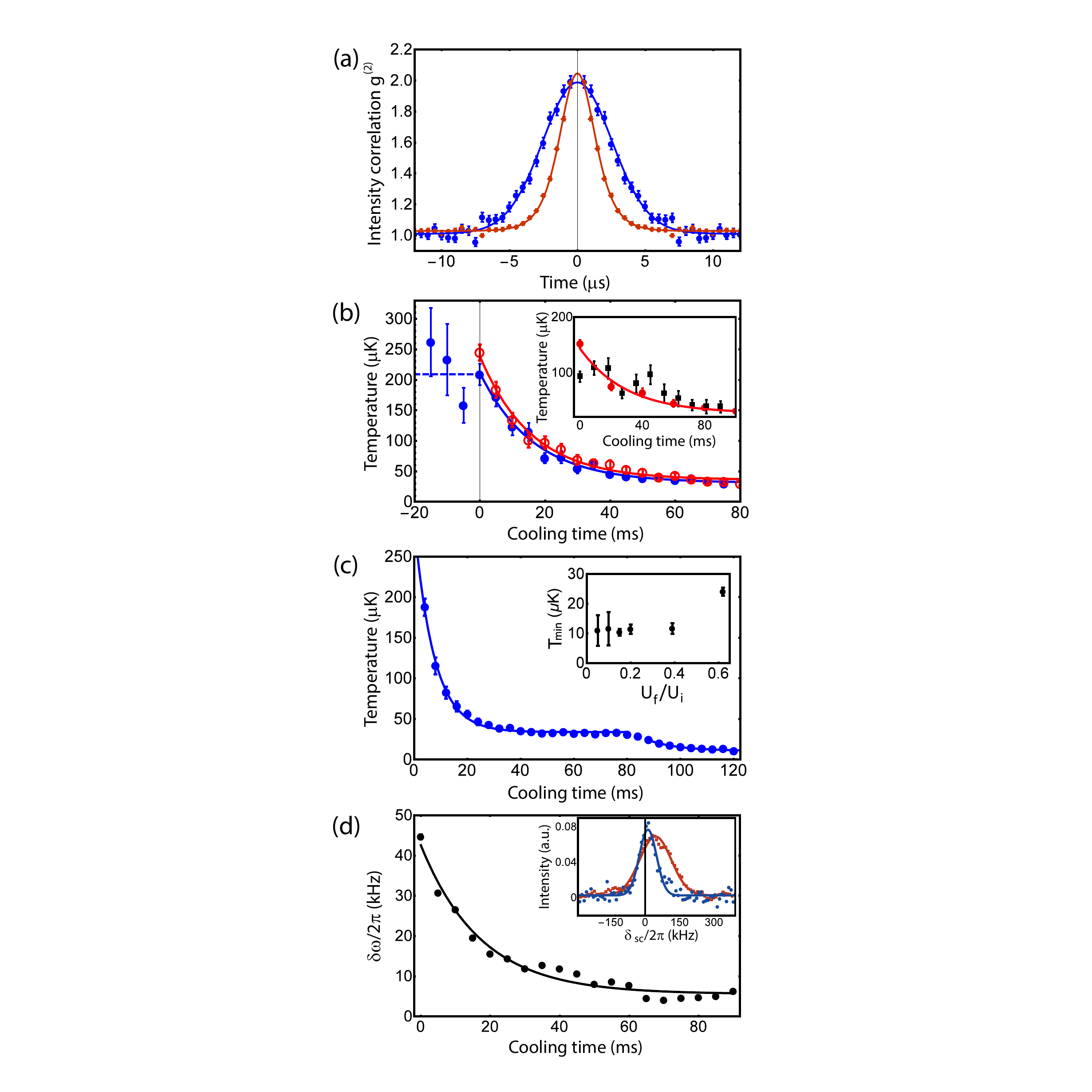}}
\caption{ (a) Photon-photon correlation function $g^{(2)}$ of the light exiting the cavity. The width of the peak is set by the Doppler decoherence time, and is used to measure the atomic temperature during cooling.  $g^{(2)}$ of the scattered light is depicted for hot (red, $T=200~\mu$K) and cold (blue, $T=30~\mu$K) atoms. 
(b) Temperature as a function of cooling time for detunings $\Delta/2\pi=-2~ \mathrm{GHz}$ (filled circles) and $\Delta/2\pi=-4 ~\mathrm{GHz}$ (empty circles) 
from atomic resonance extracted from $g^{(2)}$. 
The cooling light is tuned away from cavity resonance at $t=0~ \mathrm{ms}$ to start the cooling. The photon scattering rate per atom into the cavity $\Gamma_{cav}=11 ~\mathrm{ms}^{-1}$ is chosen the same for both detunings. The solid lines are an exponential fits to the data with $1/e$ time of $16(1)~\mathrm{ms}$ for both detunings. 
The inset shows that the atomic temperatures along different directions (black squares for temperature along $\hat{r}$, red circles for temperature along $\hat{x}$) equilibrate within 50 ms. The data are taken at atomic detuning $\Delta/2\pi=-2~\mathrm{GHz}$.
(c)  At time $t=80 ~\mathrm{ms}$ the trap depth is reduced to 15\% of its initial depth to reduce heating by trap fluctuations. In the shallower trap, the final temperature is lower and reaches $T=10  ~\mu$K, i.e., $k_B T=1.3\hbar\kappa$, close to the fundamental limit of cavity cooling. The solid lines are exponential fits for the first ($1/e$ time of $6.7(2) ~\mathrm{ms}$) and second ($1/e$ time of $11(1) ~\mathrm{ms}$) cooling stages. The inset shows the final temperature achieved as a function of the final trap depth.
(d) Frequency shift $\delta \omega$ of the peak of the atomic emission spectrum into the cavity, confirming the cavity cooling mechanism. Inset is a typical plot for the emission spectrum at the beginning (red squares, $T=200~\mu$K) and the end (blue circles, $T=30~\mu$K) of the cooling. $\delta_{sc}$ is the detuning from the incident light. 
 Error bars in this and following figures are statistical errors ($\pm 1$ standard deviation).
}
\label{fig2}
\end{figure}

%(c) The cooling rate, $R_c$, and final temperature, $T_{f}$, (shown in the inset) extracted from $g^{(2)}$. We plot the $R_c$ as  function of initial scattering rate into the cavity and $T_{f}$ as a function of final scattering rate as these are the relevant quantities when considering the cooling saturation limit (see text). The solid line is predicted from theory using equation Eq.\ref{eq:Tvst} taking into account Doppler-induced saturation of the cooling force (see the text). The dashed line in the inset is the predicted final temperature. (c) Temperature along the $x-z$ plane (filled circles) and $r-z$ (as indicated in Fig.\ref{fig1} (a)) is measured as a function of cooling time by shining cooling light in both $y$ and $r$ directions. The fact that temperature is reduced in both planes indicated cooling in three dimensions. The solid line is an exponential fit to the data with $1/e$ time of 22(2) ms and 22(5) ms for temperatures along $r-z$ and $x-z$ planes, respectively.

 %We record the photons scattered into the cavity and use three different methods to extract the atomic temperature from this information. For hot atoms, where the Doppler width of the scattered light exceeds the cavity linewidth, the Doppler spectrum can be measured by recording the light scattered into the cavity, and transmitted through the mirrors, while varying the frequency of the input cooling beam. A fit to the spectrum is then used to extract the atomic temperature (Fig.\ref{fig1}(b)).
 The temperature of the atomic ensemble is sufficiently low to ensure that the Doppler width of the atoms is comparable to or smaller than the cavity linewidth. In this regime, the intensity autocorrelation function $g^{(2)}$ of the light emerging from the cavity reflects primarily the Doppler coherence time, and can be used to extract the temperature of the atomic ensemble after correcting for the effect of the cavity linewidth (Fig.\ref{fig2}(a)). This new method of measuring temperature is in situ and in real time, non-destructive, and can be applied to small atomic samples. (The temperature can also be measured via the spectrum of the scattered light, and the two methods agree, as detailed in the Supplemental Information.) Since the atoms are confined in the Lamb-Dicke regime along the $\hat{z}$ direction ($\omega_{ax}$ far exceeds the recoil energy $E_{rec}/\hbar=2\pi \times 2$~kHz), the Doppler coherence time is set only by the temperature along the direction $\hat{x}$ of the incident cooling beam. However, all directions thermalize quickly due to interatomic collisions on a typical timescale of 15 ms, calculated from the measured Cs elastic cross section~\cite{Chin:PRL2000}.   We verify the cross thermalization by briefly applying a weak laser pulse every 3 ms with the same frequency as the cooling beam along a direction that is almost perpendicular (angle 75$^\circ$) to the direction of the cooling beam (the $\hat{r}$ direction in Fig.\ref{fig1}(a)). The data (inset to Fig.\ref{fig2}(b)) show that the $\hat{r}$ and $\hat{x}$ directions thermalize on a characteristic time below the cooling time scale of $30~\mathrm{ms}$. Note that if the atoms were not thermalizing collisionally, one could apply both beams simultaneously to cool the atoms in a horizontal plane (see S.M.), while direct axial sideband cooling along the vertical direction could be accomplished by detuning the incident frequency by the vibrational splitting $\omega_{ax}$~\cite{VV:pra2001}. \par

Fig.\ref{fig2}(b) shows the time evolution of the atomic temperature during cooling. Starting by tuning the light onto cavity resonance and recording the $g^{(2)}$ function, we measure the initial temperature of the cloud to be about 200 $\mu$K. After tuning the input light frequency
 to the red of the cavity resonance (laser-cavity detuning $\delta_i/2\pi\approx -40$ kHz) at time $t=0$, cooling begins and the temperature drops exponentially with a time constant $\tau=16(1)$ ms. The ensemble reaches a minimum temperature after {\raise.17ex\hbox{$\scriptstyle\mathtt{\sim}$}}50 ms, limited by the atomic recoil and residual heating due to trap intensity fluctuations. To demonstrate that the cooling is independent of the atomic structure and depends on the light-atom detuning, $\Delta$, only through the $\Delta$-dependent atomic polarizability and associated photon scattering rate, we compare the cooling at $\Delta/2\pi=-2~\mathrm{GHz}$ and $\Delta/2\pi=-4~\mathrm{GHz}$ from the $ F=3\to F'=2$ transition. When the power is adjusted to keep the photon scattering rate the same in both cases, we observe very similar cooling performances (Fig.~\ref{fig2}(b)), indicating that with sufficient laser power, cavity cooling can be performed at arbitrary detuning from atomic resonance. For the remainder of the data we choose the detuning $\Delta/2\pi=-2~\mathrm{GHz}$.\par
 % The results obtained from the three different measurement are consistent with each other, %and final temperature of about 20 $\mu$K extracted from different spectra% plotted in Fig.\ref{fig1}(b)-(d). We note that due to the small atom number used in this experiment,  time-of-flight absorption imaging is not suited for measuring temperature. 

The evolution of the atomic temperature $T$ for the cavity cooling of individual atoms can be modeled as~\cite{VV:pra2001}
\begin{eqnarray}
\frac{dT}{dt}= -R_c \eta \Gamma_{sc} T+H_{rec}\Gamma_{sc}+h_{trap}.
\label{eq:Tvst}
\end{eqnarray}
Here the first term with $R_c=-\frac{16E_{rec}}{3\hbar \kappa}\frac{2\delta_i/\kappa}{[1+(2\delta_i/\kappa)^2]^2} $ describes the cavity cooling due to the scattering of light into the cavity that is blue detuned by $-\delta_i$ relative to the incident light, where $E_{rec}=\hbar^2k^2/2m$ is the recoil energy associated with the wavenumber $k$ of the incident light, and $\Gamma_{sc}$ is the photon scattering rate per atom into free space. Here it has been assumed that the Doppler width is less than the cavity linewidth $\kappa$, which for our parameters is fulfilled for $T$ ${\footnotesize \lesssim}$ $300 ~\mu$K; see Ref.~\cite{VV:pra2001} for the general case. The second term with $H_{rec}=\frac{4 E_{rec}}{3 k_B} \big[ 1+\frac{\eta}{1+(2\delta_i/\kappa)^2}\big]$ describes the recoil heating associated with photon scattering both into free space and into the cavity. The third term represents the background heating due to dipole trap intensity fluctuations, which is independent of cavity cooling, and has been separately measured to be $h_{trap}=(3\pm 1)~\mu$K/ms in our system.

%\begin{figure}[!th]
%\centerline{\includegraphics[width=1.\columnwidth]{figures/fig3_v6p2.eps}}
%\caption{The cooling rate $R_c$, extracted from $g^{(2)}$, as a function of total initial scattering rate into the cavity. The solid line is predicted from theory using equation Eq.\ref{eq:Tvst} taking into account Doppler-induced saturation of the cooling force (see the text). }
%\label{fig3}
%\end{figure}

%(c) Temperature along the $x-z$ plane (filled circles) and $r-z$ (as indicated in Fig.\ref{fig1} (a)) is measured as a function of cooling time by shining cooling light in both $y$ and $r$ directions. The fact that temperature is reduced in both planes indicated cooling in three dimensions. The solid line is an exponential fit to the data with $1/e$ time of 22(2) ms and 22(5) ms for temperatures along $r-z$ and $x-z$ planes, respectively.

\begin{figure}[!th]
\centerline{\includegraphics[width=.57\columnwidth]{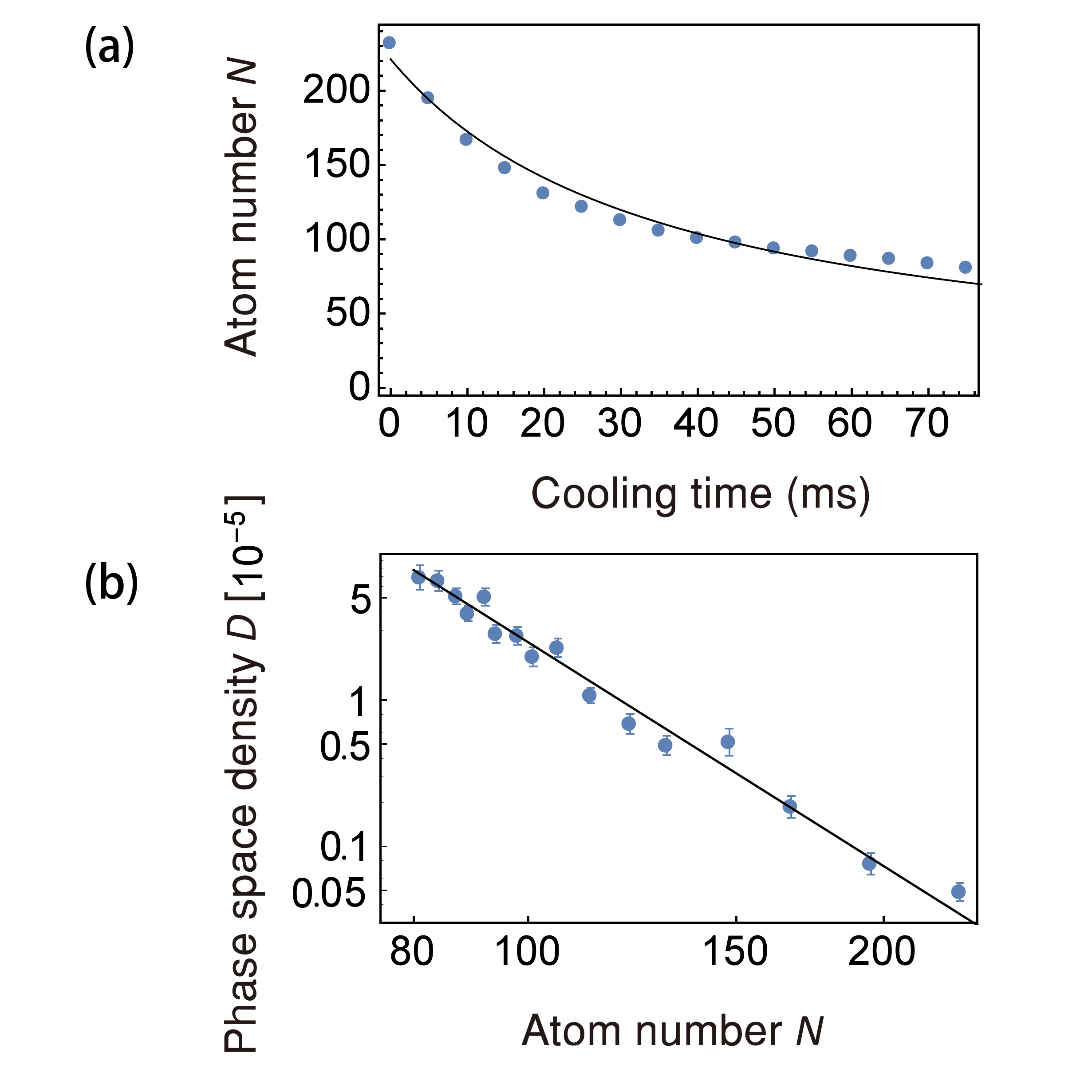}}
\caption{(a) Atom number extracted from the total scattering rate into the cavity. The solid line is a two-body loss model fitted to the experimental data with fitted two-body loss coefficient of $0.12(1)$ 1/s. (b) Phase space density $D$ as a function of remaining atom number. A linear fit between $\ln(D)$ and $\ln(N)$ is plotted as a solid line. The fit gives $\gamma=5.0(3)$, indicating very efficient cooling in spite of some light-induced loss.}
\label{fig4}
\end{figure}

The final temperature $T_{f}$ after cooling can be obtained by solving Eq.\ref{eq:Tvst} in steady state. In the limit of low trap heating and high cooperativity, the minimum temperature is reached at the cavity detuning $\delta_i=-\kappa/2$, and the final temperature is 
$T_{f}=\frac{1}{k_B}\frac{\hbar\kappa}{2} (1+\frac{2}{\eta})+\frac{3\hbar\kappa}{4E_{rec}\eta\Gamma_{sc}}h_{trap}$~\cite{VV:pra2001}. With limited cooperativity, the minimum temperature is achieved when the laser is tuned closer to the cavity resonance. For our parameters, we find a cavity detuning $\delta_i/2\pi$ around $-40$~kHz to be optimal, yielding a final temperature of $T_f=33(5)~\mu$K, in agreement with the predicted value of $30~\mu$K from $h_{trap}=2~\mu$K/ms. 

% At the cavity detuning $\delta_i=-\frac{\kappa}{2}$, the final temperature after cooling can be obtained by solving the above equation in steady state:

%\begin{eqnarray}
%T_{f}=\frac{1}{k_B}\frac{\hbar\kappa}{2} (1+\frac{2}{\eta})+\frac{3\hbar\kappa}{8E_{rec}\Gamma_{sc}}h_{trap}
%\end{eqnarray}

To verify that we can approach the theoretical limit of $T_{min}=\frac{1}{k_B}\frac{\hbar\kappa}{2} (1+\frac{2}{\eta})$ in the absence of trap heating, we reduce the trap depth $U$, which reduces trap heating. As Fig.~\ref{fig2}(c) shows, we then observe further cavity cooling down to $10(1) ~\mu$K when the trap depth $U_f$ is reduced to 15\% of its original value $U_i$. This is close to the predicted theoretical limit of $T_{min}=7 ~\mu$K for ideal cavity cooling.\par

We also verify directly that the postulated mechanism for cavity cooling, the blue shift of the cavity-scattered light relative to the incident light, is indeed responsible for the observed cooling. By interfering the light emerging from the cavity with a local oscillator detuned by 2 MHz from the frequency of the input light, we can directly monitor the emission spectrum by the atoms into the cavity at different times during the cooling sequence (Fig.\ref{fig2}(d)). The observed initial average blue shift of the cavity emission spectrum relative to the incident light of $\delta\omega/2\pi=45$ kHz in combination with the observed single-atom photon scattering rate into the cavity of $\Gamma_{cav}=$ 6 ms$^{-1}$ then predicts a cooling rate constant of $\tau=\frac{3}{2}k_B T/\hbar \delta \omega \Gamma_{cav}=24$ ms,  close to the observed value $\tau_c=17(2)$ ms.

While the temperature of the ensemble decreases, we observe some loss of atoms as a result of light-induced collisions~\cite{burnett1996laser}. The atom number, determined from the observed scattering rate into the cavity, as a function of cooling time is plotted in Fig.\ref{fig4}(a). The loss is reasonably well described by the model for light-induced collisions~\cite{burnett1996laser} $\dot{N}=-\mathcal{L} \Gamma_{sc} n \lambdabar^3 N$, where $n=n_0/2^{3/2}=1.6\times 10^{12}$ cm$^{-3}$ is the average density, $\lambdabar=k^{-1}$ the reduced probe wavelength, and $\mathcal{L}=0.76$ a parameter of order unity. To quantify the cooling efficiency in the presence of loss, we consider the logarithmic derivative $\gamma=-\mathrm{d} \ln (D)/\mathrm{d} \ln (N)$ that is used in evaporative cooling processes to characterize the cooling efficiency. Here $D=n_0\lambda_T^3$, with the peak atomic density $n_0$ and thermal de Broglie wavelength $\lambda_T$, is the  peak phase space density, and $N$ is the atom number. During 80 ms cooling time, the phase space density ramps up by over two orders while one third of the atoms remain (Fig.\ref{fig4}(b)). A fit to Fig.\ref{fig4}(b) gives $\gamma=5.0(3)$ whereas $\gamma=4$ is the largest value that has been realized in evaporative cooling~\cite{evaporative}. Furthermore, the light-induced loss could be suppressed by more than an order of magnitude by means of magnetically tuning the scattering length~\cite{vuletic1999suppression} or choosing an optimal detuning~\cite{burnett1996laser}. This indicates that cavity cooling is potentially an efficient method for increasing the phase space density. When we use circularly polarized cooling light to also optically pump the atoms into the magnetic sublevel $F=3, m_F=3$, we reach a phase space density of $D=2(1)\times 10^{-4}$, limited primarily by the cavity linewidth. \par

 In conclusion, we have demonstrated cavity cooling of the single-particle degrees of freedom of an atomic ensemble trapped inside a high-finesse optical resonator. The results could be extended in several directions. By increasing the available laser power from 3 $\mu$W to 1 W, the detuning of the light from the transition could be increased from 4 GHz to 2 THz, comparable to typical vibrational frequency splittings in molecules, and much larger than the rotational energy splittings. Working at such large detuning makes it possible to cool different molecular ro-vibrational states at the same time. Also, due to the enhancement of cavity scattering over free-space scattering by the cooperativity $\eta$, for a state-of-the-art cavity with $\eta=200$, the cooling could be faster than the optical pumping into a different molecular state. In combination with some vibrational cooling~\cite{viteau2008optical} and a magneto-optical trap for molecules~\cite{barry2014magneto}, this could allow the simultaneous cooling of molecules in many different ro-vibrational states.\par

\begin{acknowledgments}
This work was supported by the NSF, the NSF Center for Ultracold Atoms, MURI grants through AFOSR, ARO, ONR, and NASA.  Y.-T. C. acknowledges support from the Top University Strategic Alliance Fellowship.
\end{acknowledgments}
\clearpage
\bibliography{Ref}

\pagebreak
\widetext

%%%%%%%%%% Merge with supplemental materials %%%%%%%%%%
%%%%%%%%%% Prefix a "S" to all equations, figures, tables and reset the counter %%%%%%%%%%
\setcounter{equation}{0}
\setcounter{figure}{0}
\setcounter{table}{0}

\makeatletter
\renewcommand{\theequation}{S\arabic{equation}}
\renewcommand{\thefigure}{S\arabic{figure}}
%\renewcommand{\bibnumfmt}[1]{[S#1]}
%\renewcommand{\citenumfont}[1]{S#1}
%%%%%%%%%% Prefix a "S" to all equations, figures, tables and reset the counter %%%%%%%%%%

\section*{Supplemental Material}

\subsection*{Rate equation} 

The cooling force generated by the light scattered into the cavity mode is given by~\cite{VV:pra2001}:
\begin{eqnarray}
{\bf F_c}=\hbar  {\bf k}\Gamma_{cav}\frac{4 \delta_i {\bf k} \cdot{\bf v}}{(\kappa/2)^2+\delta_{sc}^2},
\end{eqnarray}
where 
\begin{eqnarray}
\delta_{sc}=\delta_i-{\bf k}\cdot{\bf v}
\end{eqnarray}
is the detuning of the scattered light relative to the cavity resonant frequency, ${\bf k} =k \hat{x}$ is the wavevector of the incident light, and $\Gamma_{cav}$ is the single-atom scattering rate into the cavity. The enhancement of the cavity scattering rate over free-space scattering rate is $\Gamma_{cav}/\Gamma_{sc}=\eta/(1+(2\delta_i/\kappa)^2)$. Since the atoms are confined in the Lamb-Dicke regime along the cavity ($\hat{z}$) direction, only the force in the $\hat{x}$ direction is responsible for the cavity cooling process. 

From the cooling force, we can calculate the rate at which thermal energy is removed from one atom:
\begin{eqnarray}
\frac{d}{dt} W= ({\bf F_c} \cdot {\bf v})\Gamma_{cav}+2 E_{rec} \Bigg[1+\frac{\eta}{1+(2\delta_i/\kappa)^2}\Bigg]\Gamma_{sc}+H_{trap},
\end{eqnarray}
where $W=\frac{3}{2}k_B T$ is the thermal energy of individual atoms. The rate equation can be written as:

\begin{eqnarray}
\frac{d}{dt} \frac{3}{2} k_B T= - k  v\frac{4 \hbar kv}{\kappa}\frac{2\delta_i/\kappa}{[1+(2\delta_i/\kappa)^2]^2}\eta \Gamma_{sc}+\nonumber \\
2 E_{rec} \Bigg[1+\frac{\eta}{1+(2\delta_i/\kappa)^2}\Bigg]\Gamma_{sc}+H_{trap}.
\label{energyrate}
\end{eqnarray}
By redefining $H_{trap}=\frac{3}{2}k_B h_{trap}$, we arrive at Eq.~1.

\subsection*{Photon-photon correlation function} 

For the light emitted from a large number of uncorrelated emitters, there is a simple relation between the first and second-order auto-correlation functions~\cite{Loudon1983}:
\begin{equation}
g^{(2)}(\tau)=1+|g^{(1)}(\tau)| ^2,
\label{eq:g2g1}
\end{equation}

where $g^{(1)}$ is defined as
\begin{equation}
g^{(1)}(\tau)= \frac{ \langle E^*(t) E(t+\tau)\rangle }{ \langle E^*(t) E(t)\rangle }. 
\label{eq:g1}
\end{equation}
 
We consider $N$ atoms with a Doppler width $\omega_D$. The electric field of the scattered light field can be written as
\begin{eqnarray}
\langle E^*(t) E(t+\tau)\rangle = E_0^2 \sum_{j=1}^N \frac{\exp{ i(\omega_0 + k v_{xj}) \tau}}{1+\big[ 2(\delta_i+ k v_{xj})/\kappa \big]^2} \nonumber  \\ 
=N E_0^2 \int_{-\infty}^{\infty} \frac{\exp{ i (\omega_0 + k v_{x}) \tau}}{1+\big[ 2(\delta_i+ k v_{x})/\kappa \big]^2}\exp{\Big( -\frac{k^2 v_x^2}{2\omega_D^2} \Big)}dv_x,
\label{eq:ee}
\end{eqnarray}
where $E_0$ is the electric field amplitude scattered from an atom. The correlation function $g^{(2)}$ of the light scattered into the cavity mode is fit with Eqs.~\ref{eq:g2g1}-\ref{eq:ee} to obtain the Doppler width. The relation between Doppler width and temperature in one dimension $(\hat{x})$, $\omega_D=k \sqrt{\frac{k_B T}{M}}$, is used to extract the temperature of the atoms.\par

\subsection*{Heterodyne measurement}

We interfere the emerging light from the cavity with a local oscillator 2 MHz detuned from the incident light on a $50/50$ beam splitter. The output light from two ports of the beamsplitter is collected using SPCM-AQRH Single Photon Counting Modules from Excelitas Technologies, and its Fourier
 transform is calculated to extract the power spectrum of the light exiting the cavity. The average frequency shift $\delta \omega$ of the scattered light into the cavity relative to the incident light is a direct evidence of cavity cooling. Thermal energy is removed from individual atoms at a rate of $\Gamma_{cav} \hbar \delta \omega$, which equals the first term in the right hand side of Eq.~\ref{energyrate}. The linewidth of the spectrum is a product of the Doppler emission spectrum of the atoms $I(\omega)=I_0 \exp{\big[-\frac{(\omega-\omega_0)^2}{2\omega_D^2}\big]}$ and the cavity transmission profile $\mathcal{T(\omega)}=\frac{1}{1+[2(\omega-\omega_c)/\kappa]^2}$, where $\omega_0$ is the frequency of the incident light, $\omega_c$ is the cavity resonance frequency, and $\omega_D$ is the Doppler width of the atomic ensemble. The temperature extracted from the emission spectrum of the cavity scattering light agrees well with that obtained from the $g^{(2)}$ function.
 
 %The spectral linewidth of the beatnote signal $\Delta \omega_{BN}$ is another way to measure the temperature while the shift of the beatnote signal $\delta \omega_{BN}$ due to atoms provides information about the cooling rate. The linewidth $\Delta \omega_{BN}$ is the product of Doppler profile of the atoms $I_D(\omega)$ and the cavity shape $\mathcal{L}(\omega-\delta_i)$, where $\delta_i$ is detuning of incident light relative to the cavity resonant frequency, 
 
 %while the shift $\delta \omega_{BN}$ gives a cooling rate due to cavity scattering of $\frac{2\hbar \delta \omega_{BN}}{3 k_B} \Gamma_{sc}$, which is the first term on the right hand side of Eq.\ref{eq:Tvst}. 
 %The data for  $\Delta \omega_{BN}$ and  $\delta \omega_{BN}$ is shown in Fig.\ref{shift}. At high temperature the beatnote width is dominated by the cavity line width (160kHz) and is not representative of the atomic temperature.

%\begin{figure*}[!th]
%\includegraphics[width=\columnwidth]{figures/SM/shiftwidth.eps}
%\caption{Linewidth and frequency shift of the interference beatnote signal as a function of cooling time.}
%\label{shift}
%\end{figure*}

\subsection*{3D cooling with two cooling beams} 

\begin{figure*}[!th]
\includegraphics[width=.4\columnwidth]{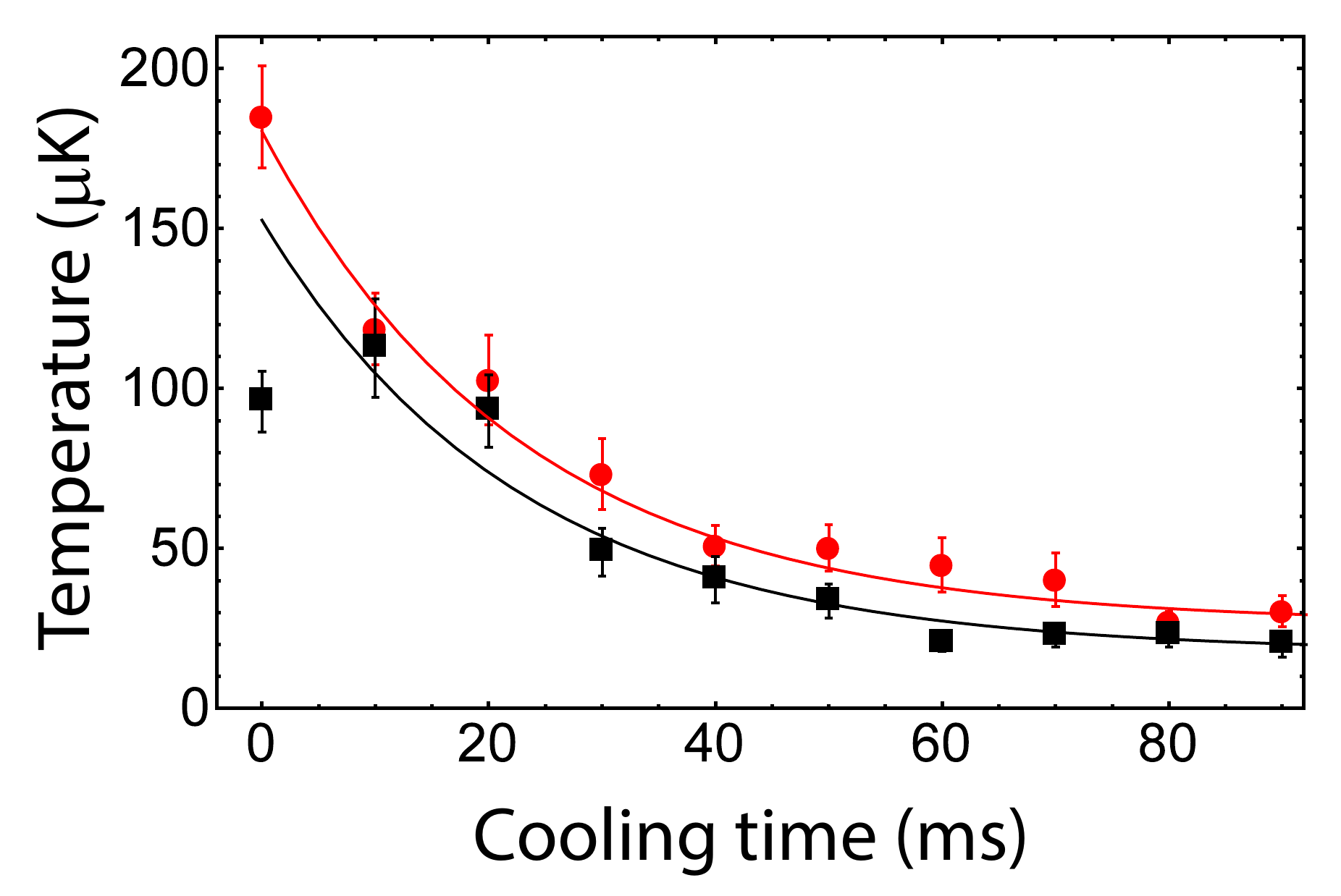}
\caption{Temperature in the $\hat{x}$ (red circles) and the $\hat{r}$ direction (black squares) as a function of cooling time when cooling beams are applied along both directions simultaneously.}
\label{Txy}
\end{figure*}

%As stated in the main text, the 3D cooling with a single cooling beam can be achieved if thermalization between different motional modes of the atoms mix faster than the cooling rate. This has been observed by cooling the atoms using a laser beam propagating along $\hat{x}$ direction and as well as shining a secondary weak laser beam pulsed along the $\hat{r}$ direction. In this case the main cooling is done by the $\hat{x}$-beam that is continuously on and the secondary  beam pulsed for 3 ms with duty cycle of 9 ms only monitor the temperature in the $r-z$ plane. The result shown in Fig.\ref{Txy} indicates that indeed the temperature is reduced in both plane as time progresses. We note that as the beam along $\hat{r}$ direction is mostly $\pi$-polarized the scattered light polarization from this beam has orthogonal polarization relative to the main cooling beam (with linear input polarization). This allow us to measure temperature in two planes independently by analyzing different polarizations separately.  
As discussed in the main text, in the weak-confinement regime, one cooling beam along the $\hat x$ direction is generating a cooling force in the x-z plane, thereby realizing 2D cooling. Any second cooling beam not parallel to the $\hat x$  axis will thus generate 3D cooling. Here we present the data when we simultaneously and continuously send light
from the $\hat x$ and $\hat r$ direction and monitor the photon-photon correlation functions separately to extract the atomic temperatures in the two directions. 
The result shown in Fig.\ref{Txy} shows a similar temperature reduction in both directions.

\end{document}